\documentclass[12pt,preprint]{aastex}
\usepackage{graphicx}

\newcommand{\sak}{V4334\,Sgr}
\slugcomment{To appear in ...}

\shorttitle{Newly ionized matter around \sak}
\shortauthors{Kerber et al.}

\begin{document}


\title{Freshly ionized matter around the final Helium shell flash object
V4334\,Sgr (Sakurai's object)\footnotemark[1]\footnotetext[1]{Based on observations 
collected at the European Southern Observatory, Chile, Proposal 67.D-0405}
}

\author{F. Kerber and N. Pirzkal}
\affil{Space Telescope European Coordinating Facility, 
	Karl-Schwarzschild-Str.\,2, D-85748, Garching, Germany}
\email{fkerber@eso.org, npirzkal@eso.org}

\author{Orsola De Marco}
\affil{American Museum of Natural History, New York, NY 10024, USA}
\email{orsola@amnh.org}

\author{M. Asplund}
\affil{Research School of Astronomy and Astrophysics, 
Mt. Stromlo Observatory, Cotter Road, Weston, ACT 2611, Australia}
\email{martin@mso.anu.edu.au}


\author{G. C. Clayton}
\affil{Louisiana State University, Baton Rouge, LA 70803, USA}
\email{gclayton@fenway.phys.lsu.edu}
\and

\author{M. R. Rosa\altaffilmark{2}}
\affil{Space Telescope European Coordinating Facility, 
	Karl-Schwarzschild-Str.\,2, D-85748, Garching, Germany}
\email{mrosa@eso.org}

\altaffiltext{2}{affiliated to Astrophysics Division of the Space Science 
Department of the European Space Agency}

\begin{abstract}
We report on the discovery of recently ionized hydrogen-deficient gas 
in the immediate circumstellar environment of the final helium shell 
flash star V4334\,Sgr (Sakurai's object). On spectra obtained with FORS2 
multi-object spectroscopy we have found spatially extended ($\sim$2\arcsec) 
emission from [\ion{N}{2}], [\ion{O}{1}], [\ion{O}{2}] 
and very faint H{$\alpha$} and [\ion{S}{2}].
In the [\ion{N}{2}] ($\lambda$$\lambda$6548,83) lines we have identified two 
components located at velocities $-350 \pm 50$ and $+200 \pm 
50$\,km\,s$^{-1}$, relative to V4334\,Sgr itself.
The full width of the [\ion{N}{2}] $\lambda$6583 feature at zero intensity
corresponds to a velocity spread of $\sim$1500\,km\,s$^{-1}$.
Based on the available data it is not possible to conclusively determine
the mechanism of ionization. Both photo-ionization, from a rapidly evolving 
central star, and shock excitation, as the result of the collision of the fast 
ouflows with slower circumstellar matter, could account for the observed lines.
The central star is still hidden behind strong dust 
absorption, since only a faint highly reddened continuum is apparent in the 
spectra.
Theory states that it will become hotter and will retrace its 
post-asymptotic giant branch evolution towards the planetary nebula domain.
Our detection of the ionized ejecta from the very late helium shell flash 
marks the beginning of a new phase in this star's amazingly rapid evolution.
\end{abstract}

\keywords{circumstellar matter --- stars: AGB and Post-AGB --- 
stars: evolution --- stars: mass loss --- 
stars: winds, outflows --- planetary nebulae: individual (V4334\,Sgr)}

\section{Introduction}

The discovery of \objectname[]{Sakurai's object} (\objectname[]{V4334 Sgr}) 
in 1996, (see \citet{due96,Kerber01} for details)
has revived strong 
interest in late phases of intermediate mass stellar evolution 
(1~M$_\odot$$\la$$M_{\rm MS}$ $\la$10~M$_\odot$) and in the physics of 
the final helium shell flash. While a rare observational event due to its 
brief 
visibility -- V4334\,Sgr is the first example since V605\,Aql 
(the central star 
of planetary nebula [PN] \objectname[]{A\,58}) 
in 1919 -- about 20\,\% of all low mass stars are expected to 
undergo a helium shell flash after they have left 
the asymptotic giant branch (AGB; Iben et al. 1983). As a result of 
the final helium shell flash the star will balloon to 
gigantic proportions while cooling and thereby returning to the AGB 
(``born-again'' giant).
From there it will retrace its own post-AGB evolution for a second time
\citep{ib96}. This will lead also to the formation of a 
second PN composed of hydrogen-poor gas and dust \citep{har96}. 
The prototypical born-again PNe are \objectname[]{A\,30} and
\objectname[]{A\,78} \citep{ja:fo,haz}, each of which have two
spatially separated and chemically distinct
nebulae, which give testimony to a helium shell flash that occurred
a few thousand years ago.

Sakurai's object, of course, is the best studied case and the only one 
observed at wavelengths other than optical. 
Its evolution suggests the following sequence of events: V4334\,Sgr
was a very hot ($>$\,100\,000\,K) central star of a PN 
\citep{Kerber99a,Pollacco99}, already 
evolving towards the white dwarf domain, when a very late helium 
shell flash occurred.
This rejuvenated V4334\,Sgr into a born--again giant which quickly 
returned to the AGB.
There, its spectrum displayed fast and massive changes as carbon bearing 
molecules formed more than a year after the 
actual helium shell flash \citep{Asplund97}, very much comparable 
to the case of V605\,Aql \citep{cl:dm97}.
While cooling, the central
star became hydrogen-deficient \citep{Asplund97,asp99}.
Following the discovery of a near-IR excess, we used the Infrared Space 
Observatory (ISO) 
to monitor the evolution of Sakurai's object
for a full year from February 1997 to February 1998. During this period the 
flux 
increased by a factor of ten in the wavelength range 4-15\,$\mu$m. 
We concluded that this was the signature of very hot dust that had recently 
formed in an extended, optically thin shell around the star 
\citep{Kerber99b}.
Shortly thereafter V4334\,Sgr showed pronounced fadings of several 
magnitudes in the visual brightness
as the dust turned optically thick in our line of sight. R\,CrB stars, truly 
irregular variables, show a very similar behavior \citep{Clayton96}.
In 1999, V4334\,Sgr became very faint in the visible ($>$ 20\,mag), following 
continued mass-loss, and has remained ``buried'' in its own 
dust ever since.
Recent evolutionary calculations \citep{Herwig01}
suggest that the star will not remain a cool giant for very long.
The resulting hardening of the central star's 
radiation field will have a fundamental 
effect on the circumstellar material and is expected to lead to the 
destruction of the 
dusty cocoon. V605 Aql, 80 years after its helium shell flash, is a
$>$\,50\,000\,K central star, working its way through a still very thick 
dust cocoon. 
The evolution of Sakurai's object again is extremely fast, since we have 
found it to be already forming a proto-PN.

\section{Observations and Data Reduction}

We observed V4334\,Sgr in service mode
on the nights of the $17^{th}$ and 
$22^{nd}$ of June, 2001, using  FORS2 (with a 2048\,$\times$\,2048 
SiTe detector) mounted on the VLT UT4 Yepun in multi-object mask mode (MXU) 
with the G300V grism, which gave us a wavelength range of 4500 to 
8000\,{\AA} at a resolution of 2.66\,{\AA}/pixel; spatial resolution was 
0$\farcs$20/pixel. The 
integration time was 45 minutes on both 
nights and the seeing was 0$\farcs$72 and 0$\farcs$88, respectively.
A 1$\arcsec$\,$\times$\,4$\arcsec$ slit was placed on the object  
at PA=45$\degr$
to match the orientation of the object in the 
field as deduced from the preparatory R- and I-band 
imaging.
Observations of the target field through the slit mask but 
without using the disperser were used to accurately locate the 
position and extent of the slits on the chip and the position of the 
object within the slits.
The data were bias-subtracted, and flat-fielded in the usual 
manner using the available nightly darks and flat-field exposures. 
Dispersion solutions for each slit and for each 
night were computed using calibration lamp spectra obtained 
through the same slit masks. Absolute flux calibration was 
achieved using the observations of the spectroscopic standard stars LTT\,9239 
and Feige\,110.
The spectra were extracted using aXe, the slitless 
spectroscopy extraction software developed at ST-ECF \citep{pir}, which
allows to extract spectra following specific directions, tilted with respect 
to the dispersion direction.

The slits we used to observe V4334\,Sgr extended further than the 
object and we used 7 pixel--wide apertures to extract an object+sky spectrum
and a spectrum of the night sky.
The latter were positioned two pixels above the aperture centered on the object
and shifted leftward by 7 pixel to account for 
the change in wavelength zero-point caused by the 45{\degr} position 
angle of our slits.  
We used the 5577.35~{\AA} atmospheric lines to accurately 
shift and match our object+sky and sky spectra. The night sky spectra were 
then subtracted from the object+sky spectra to produce 
wavelength and flux-calibrated spectra of V4334\,Sgr.
We estimate the 
error in the absolute wavelength calibration of our spectra to 
be about 0.5\,{\AA} (0.2 pixels), which corresponds to about 
20\,km\,s$^{-1}$ at 6560\,{\AA}.

\section{Results and Discussion}

\begin{figure}
\includegraphics[angle=-90,width=16.0cm]{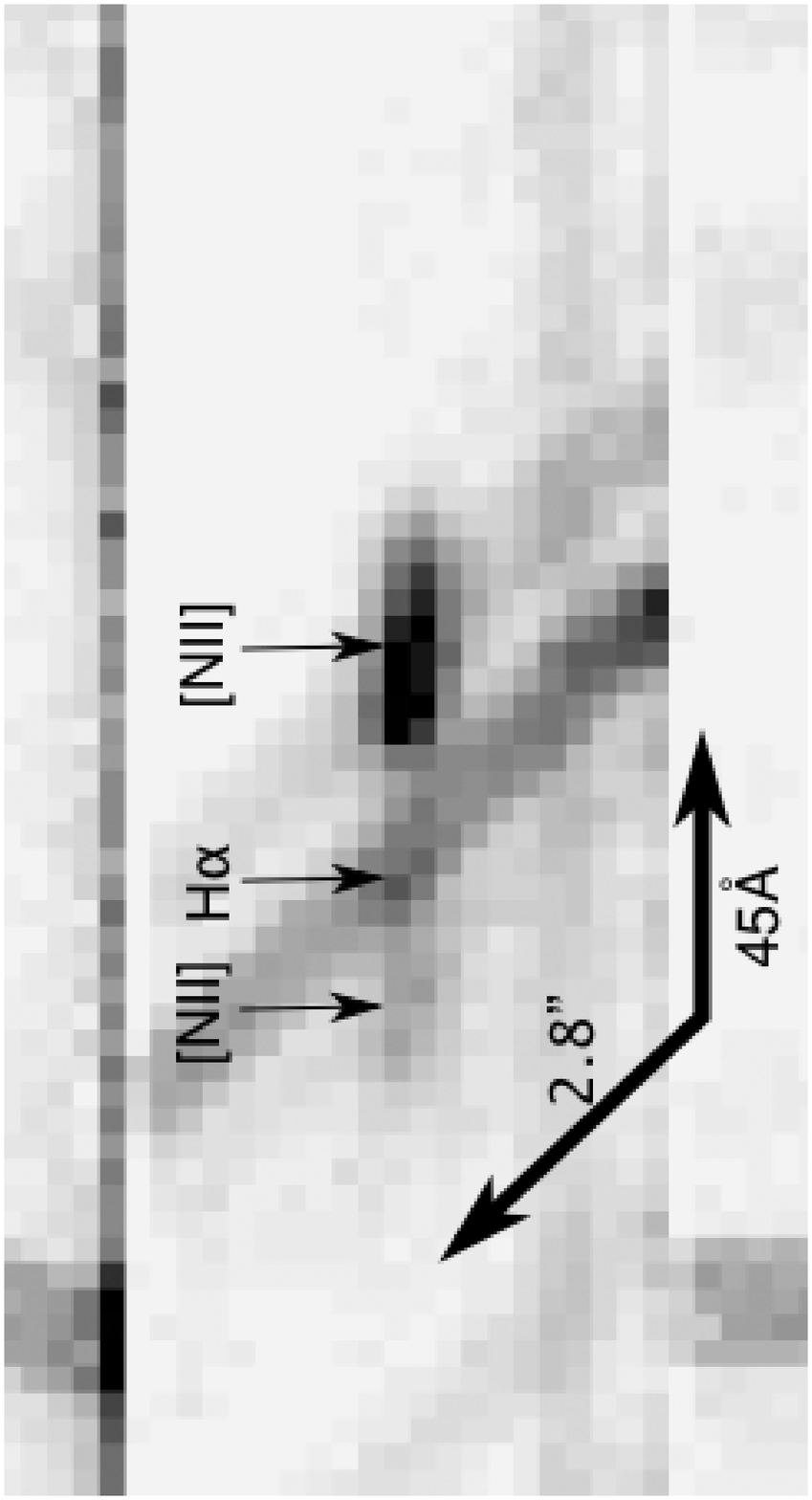}
\caption{2-D spectrum of the H$\alpha$ spectral region in V4334\,Sgr. 
The prominent feature is the [\ion{N}{2}] $\lambda$6583 line which is 
both spatially extended as well as much broader than the counterpart line 
from the old, extended PN. (Note: while the spectral direction is horizontal,
the spatial direction is, unusually, tilted at 45$\degr$ to the vertical,
direction.)}
\label{2dspec}
\end{figure}

\begin{figure}
\epsscale{0.5}
\plotone{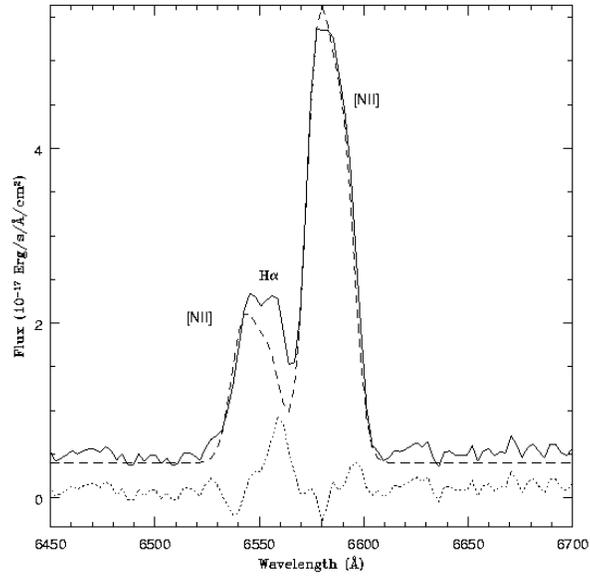}
\caption{Spectrum of the core of V4334\,Sgr around 6560\,{\AA} (solid 
line). Over-plotted is a model of the [\ion{N}{2}] doublet (dashed line)
and the observed 
spectrum after subtraction of the two [\ion{N}{2}] lines (dotted 
line). The H{$\alpha$} emission is found to be $\sim$10\,\% of the  
[\ion{N}{2}] line at 6583~{\AA}.}
\label{1dspec}
\end{figure}

We have detected spatially extended (FWHM 1$\farcs$4 in 
Fig.\,\ref{2dspec}) emission from
[\ion{N}{2}] ($\lambda$$\lambda$6548,83 and $\lambda$5755) 
as well as from [\ion{O}{1}]
($\lambda$$\lambda$6300,63), [\ion{O}{2}] ($\lambda$7325), [\ion{S}{2}] 
($\lambda$6725), and very faint H${\alpha}$
\citep{Kerber02}. At a resolution of 2.66\,{\AA}/pix the
lines of [\ion{N}{2}] and H${\alpha}$ should be resolved in the spectrum,
but they turn out to be significantly blended (solid line in
Fig.\,\ref{1dspec}). 
We performed a fit to the 
[\ion{N}{2}] line at 6583\,{\AA} and using the IRAF task {\sc
mk1dspec} 
we created a spectrum containing the two [\ion{N}{2}] lines only 
(short dashed line), 
which we then subtracted from the observed spectrum.
The resulting spectrum (dotted
line) was used to estimate the flux of the H${\alpha}$ to be $\sim$10\,\%
of the [\ion{N}{2}] line $\lambda$6583, indicating that the new ejecta are
extremely hydrogen-deficient. 
This is in strong contrast to the chemical
composition of the much larger (diameter=44\,$\arcsec$; Jacoby et al. 1998)  
old PN, for which \citet{Kerber99a} and \citet{Pollacco99} 
showed H$\alpha$ to be more than twice as strong as
[\ion{N}{2}] $\lambda$6583.  Hydrogen-deficient ejecta are, on the other
hand, in line with earlier studies of the photospheric composition of
V4334\,Sgr during the helium shell flash \citep{asp99}.
Table~\ref{t:em} lists the observed lines with their respective
fluxes. Velocities relative to the core of V4334\,Sgr  
({$v{_{\mathrm{hel}}}$} = 104\,$\pm$\,3\,km\,s$^{-1}$; 
Kipper \& Klochkova 1997) are given
only when individual components could reliably be identified.

No proper plasma diagnostic is possible with the few lines observed. 
However, the [\ion{N}{2}]
$\lambda$$\lambda$6548,83/$\lambda$5755 line ratio
suggests $T_e$$\sim$13\,000\,K,
assuming interstellar extinction only (E(B-V) = (0.71\,$\pm$\,0.09) mag;
Pollacco 1999). This result must be
considered a rough estimate, due to the faintness of the [\ion{N}{2}] line at
5755\,{\AA} and because of additional unknown circumstellar extinction. 
Similarly, the
[\ion{S}{2}] $\lambda$6716/$\lambda$6731 ratio indicates 
$n_{\mathrm{e}}$$\la$\,1000\,cm$^{-3}$. 
No lines requiring excitation energies in
excess of 15\,eV are present, in particular we do not detect excess flux
in the [\ion{O}{3}] line at 5007\,{\AA}, beside what is attributable to the old
PN. It is possible that a weak circumstellar [\ion{O}{3}] line might be
rendered invisible by circumstellar extinction.
\placetable{t:em}
If one assumes photo-ionization only, the spectra would indicate that the 
stellar temperature
has increased from about 5000-6000\,K in 1997 \citep{asp99,pa:geb:02} to 
$\ga$\,20\,000\,K, necessary to
ionize neutral nitrogen and oxygen. As the star is predicted to become 
hotter, the [\ion{Ne}{2}] line at 12.8 $\mu$m (excitation energy 21.6\,eV)  
will be an excellent indicator to follow future temperature changes.
If the star itself is developing a Wolf-Rayet-type wind, as is likely to
happen to a heating, hydrogen-deficient star, then the heavy line
blanketing will transfer hard ionizing radiation to longer wavelengths. 
Compared to an H-rich star, this would 
delay the ionizing of circumstellar material, because the heating star 
would appear to be cooler than it actually is.

There are a number of young proto-PNe like e.g. \objectname[]{OH\,231.8+4.2}
\citep{bu:02}
or \objectname[]{LW\,LMi} \citep{sch:02} which already contain ionized
material, while their central star is still too cool to photo-ionize
hydrogen, although in some cases a hot unseen binary is suspected.
These objects also show high velocity outflows and the ionization seen is 
attributed to shocks as the fast outflows collide with slower previously 
ejected matter. Therefore shock excitation could produce ionized material
around V4334\,Sgr while it is still a cool star on the AGB.

In order to distinguish between the two ionization mechanisms we have looked 
into some existing models of shock excitation and we have found resonable 
agreement for the line ratios of [\ion{N}{2}], [\ion{O}{1}] and [\ion{O}{2}]  
with bow shock model 3 and good agreement with model 4 described by 
\citet{hart:87}. Both models have
a electron density of 300\,cm$^{-1}$. The 
velocity is 100\,km\,s$^{-1}$ for model 4 and 200\,km\,s$^{-1}$ for 
model 3. See their work for details of the models.
Significant differences in the line strengths exist for H${\alpha}$, 
which of course cannot be 
reproduced by a model using solar abundances, and more interestingly, 
for [\ion{S}{2}], which is found to be considerably fainter in V4334\,Sgr 
than in all the shock models.
Using the diagnostic diagram of \citet{sab:77}
we have further studied the mechanism of ionization in V4334\,Sgr.
The bow shock models 3 and 4 are located in the domain of supernova remnants
(SNRs) in this diagram giving testimony to their mechanism of ionization.
Without using a reddening correction for V4334\,Sgr we obtain
$log$ H${\alpha}$/[\ion{S}{2}] $\approx$ 0.3 and $log$ 
H${\alpha}$/[\ion{N}{2}] 
$\approx$ $-$1.25. This position does not coincide with any of the object 
classes in the diagram. While the large H${\alpha}$/[\ion{S}{2}] might argue 
against shock excitation, the line ratio cannot be used as a diagnostic tool 
in the H-deficient plasma. Moreover, emission from the different lines
may arise from 
physically distinct regions around V4334\,Sgr.
We therefore conclude that both photo-ionization, from a heating central star,
and shock excitation, as the result of the collision of the fast 
ouflows with slower circumstellar matter, could account for the observed 
lines. Clearly, better data and H-deficient
shock models are required to determine the relative importance of the two 
processes.

Based on the observational data we propose that
V4334\,Sgr is still hidden from sight, since
we only detect a very faint red continuum.
This is confirmed by the R magnitude of
$\approx$\,22 which is about 10 magnitudes fainter than before the onset
of dust formation (e.g., Kerber 2001). We estimate that an extinction of 
A$_{\mathrm{V}}$ $\geq$ 7\,mag is sufficient to redden the spectrum of a
20\,000~K star to the one observed. The broad forbidden emission lines
derive from ionized matter located in two extended lobes which suffer from
much less extinction, possibly not much more than the interstellar
extinction.
These lobes are the result of high velocity outflows from the central
source. Hence, V4334\,Sgr now appears to be forming a new H-deficient, bipolar
proto-PN. Since the central star is highly reddened, while the lobes suffer
less extinction, we deduce the presence of an incomplete dust shell or
torus with polar openings. We assume that the axis of the bipolar nebula 
is thus tilted by $\sim$\,45$\degr$ with respect to the line of sight: 
a smaller angle would expose the star, while a much larger one would call for 
un-physically high velocities in the ejected lobes.
Consequently the radial velocities of $-$350 and 200\,km\,s$^{-1}$ are
lower limits of the physical outflow velocities of the lobes. The
velocities observed would therefore be extremely high, a fact that is not
unknown amongst highly bipolar proto-PNe such as 
M\,2--9 \citep{doy00} and He\,3-1475 \citep{sa:sa01,bo:ha01}, 
which are shown to possess
a very fast ($\ge$ 1000\,km\,s$^{-1}$) post-AGB wind 
only a few decades after the stars left the AGB.  
Trigonometric arguments based on the time between outburst and our
observations (6.5 yr), a short distance to V4334 Sgr of 1~kpc (for a
review of the distance to V4334~Sgr problem, see Jacoby et
al. 1998), and an angular radius of the ionized nebula of 1$\arcsec$, 
result in an
inclination angle for the outflows of 64$\degr$ (assuming the higher lobe
velocity of $-$350\,km\,s$^{-1}$). Such an angle is consistent with the 
central star
being hidden behind a thick dust torus. In this scenario the true outflow
space velocity would be $\sim$800\,km\,s$^{-1}$. If we advocated a higher 
distance
to V4334 Sgr of $\sim$5~kpc, the angle would be 84$\degr$, and the space
velocity of the flow would become 3300~km\,s$^{-1}$, too large to be credible.
This simplistic
calculation therefore also points to the low distance to V4334 Sgr.

Usually a binary central star is assumed for PNe with highly bipolar
morphology. No indication of binarity has been
found for V4334\,Sgr over the past six years despite intense scrutiny. During 
the early
phases of the final helium shell flash no high velocities were observed in the
photospheric lines. Observations in 1999, though, showed a strong P\,Cygni
profile in the He~{\sc i} line at 1.083 $\mu$m \citep{ey:99,ev:geb:02}, 
indicative of velocities between 500 -- 800\,km\,s$^{-1}$. 
The high velocities found in V4334\,Sgr today -- only a few years after it 
arrived on the AGB -- could indicate that a fast wind from the central star 
can develop very quickly on the AGB. 

The closest relative to V4334\,Sgr seems to be V605\,Aql, which
experienced its helium shell flash in 1919, and exhibited
a similar light curve to that
of V4334\,Sgr between 1919 and 1923. V605 Aql was not observed for sixty
years, before being rediscovered in the 1980s \citep{sei}. The
central star has a temperature of $>$\,50\,000\,K 
\citep{cl:dm97}, but the object is still very faint visually. 
HST observations of V605\,Aql reveal a very small 
(0.7\,arcsec) inner knot in the [\ion{O}{3}] line \citep{bo93,cl:dm97}. 
Apparently this is radiation leaking from a still
largely intact cocoon of dust. The velocities observed in the core of
V605\,Aql are about 100\,km\,s$^{-1}$ \citep{pol:92}, almost
an order of magnitude smaller than for V4334\,Sgr. 
From the observational facts described above we deduce that the fundamental
difference between V605\,Aql and Sakurai's object is in the geometry
of the mass-loss. In the case of V605\,Aql one has to assume a rather
symmetric distribution of circumstellar matter. Therefore the central 
star's hard
radiation field is ionizing the matter around it but it has to work its
way through the circumstellar dust shell. In the case of Sakurai's object
the data suggest that the mass-loss was highly non-spherical and that high
velocity outflows of ionized matter in bipolar lobes have already reached
significant distances from the central star, where it can be observed with
little interference from circumstellar extinction.

A geometry similar to the one described for V4334\,Sgr has been found
in A\,30, whose inner, H--deficient PN is well extended due to its expansion,
since the helium shell flash, which occurred a few thousand years ago. 
HST WFPC\,II observations revealed a highly complex structure of the 
ejecta \citep{bo:ha2} as a result of distinctly 
non-spherical mass loss. In particular the distribution of the knotty ejecta
indicates the former presence of an equatorial disk, which has been broken down
in individual clumps by the fast ($>$\,1000\,km\,s$^{-1}$) wind from the hot 
CS, and 
there are also knots at highly symmetrical locations in the polar directions.

\section{Conclusions}
We have discovered freshly ionized H-deficient
matter around the core of V4334\,Sgr.
Currently it is not possible to conclusively determine
the mechanism of ionization: photo-ionization or collisional ionization.
The forbidden emission lines are produced in high velocity outflows, with
components at velocities of $-$350 and +200\,km\,s$^{-1}$ relative to 
V4334\,Sgr itself.
The total width of the [\ion{N}{2}] $\lambda$6583 feature corresponds to a 
dispersion of about 1500\,km\,s$^{-1}$. A detailed discussion of the velocity 
structure of V4334\,Sgr will be given in a subsequent paper.
Using all available information we conclude that
the central star is still hidden behind a thick torus of dust, whereas
ionized matter is located in two extended lobes of high velocity indicating 
that Sakurai's object is forming a bipolar proto-PN. 
This difference in geometry makes it much 
easier to follow the further evolution of V4334\,Sgr than for V605\,Aql 
which is still hidden inside a largely complete dust cocoon more than 80 years
after it experienced its final
helium shell flash. 
It will be important to verify the evolutionary scenario adopted,
by observing the appearance of the [\ion{Ne}{2}] line at 12.8\,$\mu$m, which 
will give 
direct evidence of an increase of stellar temperature.
Furthermore adaptive optics at 8\,m class telescopes will enable us to 
spatially resolve the structure of V4334\,Sgr and follow the evolution of 
the circumstellar matter.



\acknowledgments

O. D. acknowledges financial support from Janet Jeppson Asimov.\newline
We thank the referee for his valuable suggestions to study the relevance of 
shock excitation as a viable explanation of the ionized matter.

\clearpage

\begin{deluxetable}{cccc}
\tabletypesize{\scriptsize}
\tablecaption{Emission lines observed in V4334\,Sgr's inner nebula.\label{t:em}} 
\tablewidth{0pt}
\tablehead{
Species        & $\lambda_{\mathrm{lab}}$  & 
Velocity       & Flux\\
               & ({\AA}) & 
(km\,s$^{-1}$) & (10{$^{-17}$}erg cm$^{-2}$ s$^{-1}$)
}
\startdata
{[{N}~{\footnotesize{\sc i}}]}   & 5197.9, 5200.2 & \nodata  & 8	\\
{[{N}~{\footnotesize{\sc ii}}]}   & 5754.6  & $-$420 &	2\\
{[{O}~{\footnotesize{\sc i}}]}   & 6300.3 & $-$315/190  & 30\\
{[{O}~{\footnotesize{\sc i}}]}   & 6363.8 & $-$245 & 8	\\
{[{N}~{\footnotesize{\sc ii}}]}   & 6548.0 & \nodata & 22/18	\\
H$\alpha$   & 6562.8 & $-$253 & 9	\\
{[{N}~{\footnotesize{\sc ii}}]}   & 6583.5 & $-$347/204 & 65.5/54.5	\\
{[{S}~{\footnotesize{\sc ii}}]}	& 6716.4 & \nodata & 2.5 \\
{[{S}~{\footnotesize{\sc ii}}]}	& 6730.8 & \nodata & 2 \\
{[{O}~{\footnotesize{\sc ii}}]}   &  7319, 7330 & \nodata & 62	\\
%
 \enddata

\end{deluxetable}


\begin{thebibliography}{}
   \bibitem[Asplund et al.(1997)]{Asplund97}Asplund, M., Gustafsson, B., 
Lambert, D.L., Kameswara Rao, N. 1997, \aap, 321, L17
\bibitem[Asplund et al.(1999)]{asp99}
	 Asplund, M., Lambert, D. L., Kipper, T., Pollacco, D., Shetrone, 
M. D. 1999, \aap, 343, 507 
\bibitem[Bond et al.(1993)]{bo93}
	Bond, H. E., Meakes, M. G., Renzini, A. 1993, in Planetary Nebulae, 
	IAU Symp. 155, eds. R. Weinberger and A. Acker (Kluwer, Dortrecht), 
	p.499
\bibitem[Borkowski et al.(1995)]{bo:ha2}
	Borkowski, K. J., Harrington, J. P., Tsvetanov, Z. I.
	1995, \apjl, 449, L143
\bibitem[Borkowski \& Harrington(2001)]{bo:ha01}
	Borkowski, K. J., Harrington, J. P. 2001, \apj, 550, 778
\bibitem[Bujarrabal et al.(2002)]{bu:02}
	Bujarrabal, V., Alcolea, J., S\'anchez Contreras, C., Sahai, R. 2002, \aap, 389, 271
   \bibitem[Clayton(1996)]{Clayton96}Clayton, G. C. 1996, \pasp, 108, 225
   \bibitem[Clayton \& De Marco(1997)]{cl:dm97}Clayton, G. C., De Marco, O.
 1997, \aj, 114, 2679
\bibitem[Duerbeck \& Benetti(1996)]{due96} Duerbeck, H. W., Benetti, S.
 1996, \apj, 468, L111
\bibitem[Doyle et al.(2000)]{doy00} Doyle, S., Balick, B., Corradi, R. L. M., 
Schwarz, H. E. 2000, \aj, 119, 1339
\bibitem[Evans et al.(2002)]{ev:geb:02}Evans, A., Geballe, T. R., Tyne, 
V. H., Pollacco, D., Eyres, S. P. S., Smalley, B. 2002, \mnras, 332, L69   
\bibitem[Eyres et al.(1999)]{ey:99}Eyres, S. P. S., Smalley, B.,  Geballe, 
T. R., Evans, A., Asplund, M., Tyne, V.H. 1999, \mnras, 307, L11   
\bibitem[Guerrero \& Manchado(1996)]{gu:ma:96}Guerrero, M. A., Manchado, 
A. 1996, \apj, 472, 711
\bibitem[Harrington(1996)]{har96}
	Harrington, P. J. 1996, ASP Conference Series 96, Hydrogen-Deficient 
	Stars, C.S. Jeffery and U. Heber (eds.), p.193
\bibitem[Hartigan et al.(1987)]{hart:87}
	Hartigan, P., Raymond, J., Hartmann, L. 1987, \apj, 316, 323
\bibitem[Hazard et al.(1980)]{haz}
	Hazard, C., Terlevich, R., Morton, D. C., Sargent, W. L. W., Ferland, 
	G. 1980, \nat, 285, 463
\bibitem[Herwig(2001)]{Herwig01}Herwig, F. 2001, \apjl, 554, L71
\bibitem[Iben et al.(1983)]{ib83}
	Iben, I., Kaler, J. B., Truran, J. W., Renzini, A. 1983, \apj, 265, 605
\bibitem[Iben \& MacDonald(1996)]{ib96}
	Iben, I.Jr., MacDonald, J. 1996, White Dwarfs, D. Koester, K. 
	Werner (eds.), Springer Lecture Notes in Physics, p.48
\bibitem[Jacoby \& Ford(1983)]{ja:fo}
     Jacoby, G. H., Ford, H. C. 1983, \apj, 266, 298
\bibitem[Jacoby et al.(1998)]{jac98}Jacoby, G.H., De Marco, O., Sawyer, D.G.
	1998, \aj, 116, 1367
   \bibitem[Kerber et al.(1999b)]{Kerber99b}Kerber, F.,  
Blommaert, J. A. D. L., Kimeswenger, S., Groenewegen, M. A. T., K\"aufl, H.U.,
Asplund, M. 1999b, \aap, 350, L27
   \bibitem[Kerber et al.(1999a)]{Kerber99a} Kerber F., K\"oppen, J., 
Roth, M., Trager, S. C. 1999a,  \aap, 344, L79
   \bibitem[Kerber(2001)]{Kerber01}Kerber, F. 2001, \apss, 91
\bibitem[Kerber et al.(2002)]{Kerber02} Kerber, F., Pirzkal, N., Rosa, M. R. 
2002, \iaucirc, 7879
  \bibitem[Kipper \& Klochkova(1997)]{ki:kl97}Kipper, T., Klochkova, V. G. 
1997,  \aap, 324, L65
   \bibitem[Pavlenko \& Geballe(2002)]{pa:geb:02}Pavlenko, Y. V., Geballe, 
T. R. \aap, 390, 621
   \bibitem[Pirzkal et al.(2001)]{pir} Pirzkal, N., Pasquali, A., 
Demleitner, M. 2001, ST-ECF Newsletter 29, 5
\bibitem[Pollacco et al.(1992)]{pol:92}Pollacco, D., Lawson, W. A., Clegg,
R. E. S., Hill, P. W. 1992, \mnras, 257, 33p
  \bibitem[Pollacco(1999)]{Pollacco99}Pollacco, D. 1999, \mnras, 304, 127
\bibitem[Sabbadin et al.(1977)]{sab:77}
	Sabbadin, F., Minello, S., Bianchini, A. 1977, \aap, 60, 147 
\bibitem[S\'anchez Contreras, C. \& Sahai(2001)]{sa:sa01} S\'anchez Contreras, C., Sahai, R. 2001, \apjl, 553, L173
\bibitem[Schmidt et al.(2002)]{sch:02}
	Schmidt, G. D., Hines, D. C., Swift, S. 2002, \apj, 576, 429
\bibitem[Seitter(1987)]{sei}
     Seitter, W. C. 1987, ESO Messenger 50, 14 
\end{thebibliography}
\end{document}